\documentclass[prb,aps,showpacs]{revtex4}

\usepackage{dcolumn}
\usepackage{bm}
\usepackage{amsmath}
\usepackage{graphics}
\usepackage{epsfig}

\begin{document}

\title{Nanoscale Quantum Solvation of para-H$_2$ around the Linear OCS Molecule \\ inside $^4$He Droplets}

\author{Yongkyung Kwon$^{a,b}$ and K. Birgitta Whaley$^b$}

\affiliation{$^a$Department of Physics, Konkuk University, Seoul 143-701, Korea\\
$^b$Department of Chemistry,
 University of California, Berkeley, CA 94720, USA}


\begin{abstract}
    We present a microscopic analysis of the quantum solvation structures of 
    para-H$_2$
    around the OCS molecule when embedded in low temperature $^4$He droplets.
    The structures of clusters containing $M=5$ and $6$ para-H$_2$ 
    molecules are compared with corresponding structures for $M=1$ 
    (OCS-H$_2$ complex) and $M=17$ (a full solvation shell), 
    as well as with the clusters in the absence of helium.  
    We find that the helium has negligible effect on the structures 
    for the small and large 
    OCS(H$_2$)$_M$ clusters, but that it modifies the cluster structure 
    for $M=6$.  We discuss implications of these results for the onset 
    of superfluidity in the solvating hydrogen shell and 
    for spectroscopic measurements.

\end{abstract}


\maketitle

\section{INTRODUCTION}
Finite size clusters and droplets of helium and para-hydrogen provide unique 
opportunities to find new superfluid states and to probe these with molecular
and atomic impurities. Helium clusters can be formed at temperatures of
T=$0.15 - 0.4$~K, and can be doped with a range of molecular impurities,
which are subsequently probed with a variety of spectroscopic
techniques.\cite{toennies01}
Spectroscopic features such as the observation
of free molecular rotation in $^4$He droplets but not in $^3$He 
droplets, reduced rotational constants
saturating at large droplet values already for small 
clusters, and
energy gaps to phonon wings in electronic spectra
have all been quantitatively
explained in terms of the superfluid nature of a finite $^4$He cluster
into which a molecule is doped.\cite{toennies01}

Less progress has been made for the analogous clusters of molecular H$_2$ whose
properties are also expected to be dominated by quantum effects, both zero-point
delocalization and permutation symmetry, but which are considerably 
more strongly bound than helium clusters and in which there is therefore
appreciable competition between structuring forces and 
quantum effects. 
Theoretical work has predicted finite superfluid response at low temperatures
for very small hydrogen
clusters\cite{sindzingre91}, for doped two-dimensional hydrogen 
films\cite{gordillo97}, and for hydrogen clusters solvating a small 
molecule\cite{kwon02}. This last system 
is accessible experimentally via the above-mentioned
techniques of doping helium droplets and subsequently probing
the embedded impurity species spectroscopically.  The linear OCS molecule 
has provided a very useful probe for these studies and a series of
spectroscopic experiments have recently been carried out
with OCS wrapped by from $M=1$ to $M=16$ hydrogen molecules, embedded into 
cold helium droplets.\cite{grebenev00,grebenev01,grebenev01a}

Here we present a series of theoretical studies 
of hydrogen-solvated OCS embedded inside pure $^4$He clusters,
focusing on several sizes of hydrogen solvation that provide critical
insight into the onset of molecular superfluidity for hydrogen.
We treat the para-H$_2$ molecules as spherical particles like helium
atoms and the OCS molecule as fixed and non-rotating. 
A sum of pair potentials consisting of
H$_2$-H$_2$, H$_2$-He, He-He, H$_2$-OCS, and He-OCS terms is used to represent 
the interaction potential between the three different particles involved. 
As discussed in detail in Ref.~\onlinecite{kwon03},
the OCS-H$_2$ interaction has a very similar topology
to the OCS-He potential, but a much deeper global minimum.
We employ the path-integral Monte Carlo (PIMC) method 
which allows us to make a quantitative estimate of the superfluidity
of the helium and hydrogen components.\cite{ceperley95}
The global superfluid fraction can be evaluated by an estimator
written in terms of the projected area of the Feynman paths 
(see Eq. (5) of Ref.~\onlinecite{kwon03}).

\section{OCS(H$_2$)$_M$ COMPLEXES IN $^4$He$_N$ DROPLETS}
We have previously characterized the structure and rotational spectroscopy 
of the both the $M=1$ OCS-H$_2$ complex inside a helium cluster\cite{kwon03}
and the fully 
hydrogen-solvated $M=17$ OCS(H$_2$)$_{17}$ cluster\cite{kwon02}. 
The $M=1$ complex was seen to be well described by a rigidly attached 
H$_2$ molecule resulting in an asymmetric top, while the fully solvated $M=17$ 
cluster was seen to possess an anisotropic superfluid response that implies 
a decoupling of the hydrogen shell to rotations about the OCS molecular axis.
The structure of this hydrogen solvation layer 
for $M=17$ consists of multiple annular 
bands around OCS,
while the single hydrogen molecule for $M=1$ is located in the global minimum 
of the OCS-H$_2$ (and OCS-He) potential. 
Neither the $M=1$ nor $M=17$ clusters show significant 
change in their structure on adding helium.
The superfluid response for $M=17$ was seen to be characterized 
by permutation exchanges between different annular bands.  
The structure of the intermediate size clusters is thus critical 
for understanding the onset of this nanoscale molecular superfluidity 
of para-hydrogen. 

\begin{figure}
\centerline{\includegraphics[height=2.5in]{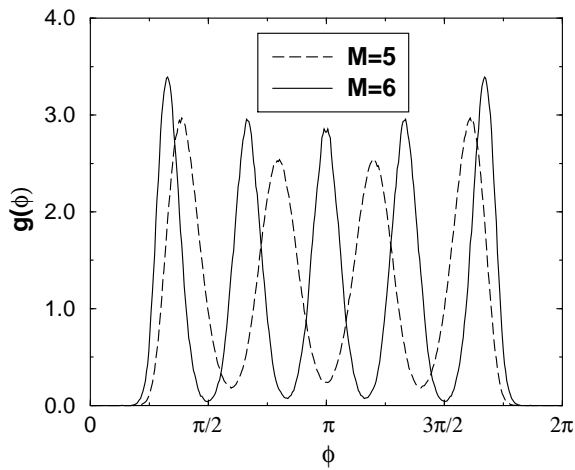}}
\caption{Azimuthal angular pair correlation function $g(\phi)$ 
in OCS(H$_2$)$_M$
complexes with $M=5$ and 6 inside a droplet of 39 $^4$He atoms at T=$0.3125$~K.}
\label{fig:gphi}
\end{figure}

As noted earlier in Ref.~\onlinecite{patel01}, 
five hydrogen molecules in a OCS(H$_2$)$_5$ complex 
inside a cluster of $^4$He$_{39}$
can form a complete band of density encircling
the linear OCS molecule, removing all helium atoms
from the waist location of lowest OCS-He/OCS-H$_2$ potential energy. 
In order to quantify the angular distribution of these hydrogen
molecules, we compute their azimuthal angular pair correlation
function.
Figure \ref{fig:gphi} shows that the five hydrogen molecules
show a very structured angular pair correlation 
that is consistent with five approximately equally-spaced particles located 
in an annular band around the OCS axis.  This can be understood
as a result of the confinement in the waist region together 
with the hard-core repulsion between hydrogen molecules. 
These structural properties of the $M=5$ complex 
differ very little from those computed 
for the free complex of the same size
without surrounding helium.

Figure 2(a) shows the hydrogen and helium density distributions
of the OCS(H$_2$)$_6$ complex inside a $^4$He$_{39}$ cluster at T=$0.3125$~K.
All of the six hydrogen molecules are distributed
at the global minimum of the OCS-H$_2$ potential and form an
annular band, just like the $M=5$ case.  
The corresponding azimuthal angular pair correlation function
is shown in Figure~\ref{fig:gphi}.  
It is evident that the six H$_2$ molecules show a more structured
angular pair correlation (solid line) than the five H$_2$ molecules 
in the OCS(H$_2$)$_5$ complex (dashed line).
This implies that the azimuthal distribution in the six-hydrogen band
is more compact than in the five-hydrogen band.
For the $M=6$ cluster there is now also 
a marked dependence of the structure on the presence of helium. 
Diffusion Monte Carlo calculations have shown that 
the hydrogen density distribution in the free OCS(H$_2$)$_6$ complex without
any surrounding helium shows a five-hydrogen band located
at the waist of the OCS, and one additional H$_2$ molecule at the oxygen side
next to the band.\cite{paesani03}  
Our PIMC calculations confirm 
that this zero-temperature structure is also 
the most stable structure at T=$0.625$~K in the absence of helium. 
Hence we may conclude that the six-hydrogen band is not stable without
the presence of additional helium atoms. 
This is understandable when considering that
the hydrogen-hydrogen distance in the six-hydrogen band located
at the global minimum of the OCS-H$_2$ potential
is shorter than the minimum energy distance ($\sim 3.47$ \AA) 
between two H$_2$ molecules.
We find that the azimuthally symmetric six-hydrogen band is broader 
in the z-direction than the five-hydrogen
band formed in the OCS(H$_2$)$_5$ complex inside the $^4$He$_{39}$ droplet.
This greater breadth in the axial direction can also be rationalized as
a result of the shorter hydrogen-hydrogen distance in the six-hydrogen complex. 
Since these H$_2$ molecules are confined in a narrow band and
effectively constitute a one-dimensional system (see Figure~\ref{fig:gphi}),
the only possible exchanges among them are the cyclic ones involving
all six molecules (five for the $M=5$ complex).
Finding a temperature regime in which
such exchange-coupled paths are realized and characterization of 
a possible superfluid/supersolid state require more computational work.

\begin{figure}
\centerline{\includegraphics[height=1.8in,width=2.8in]{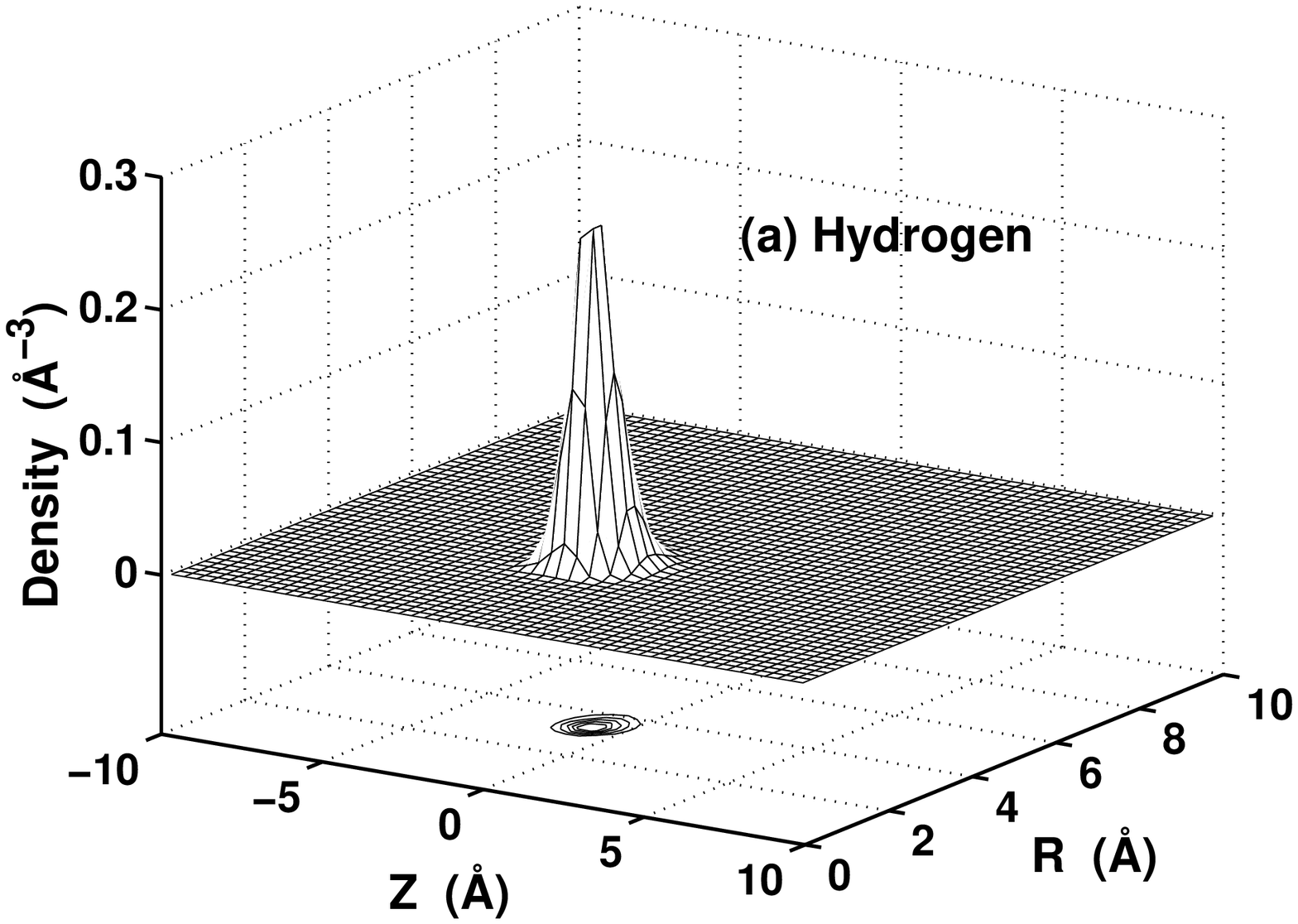}\includegraphics[height=1.8in,width=2.8in]{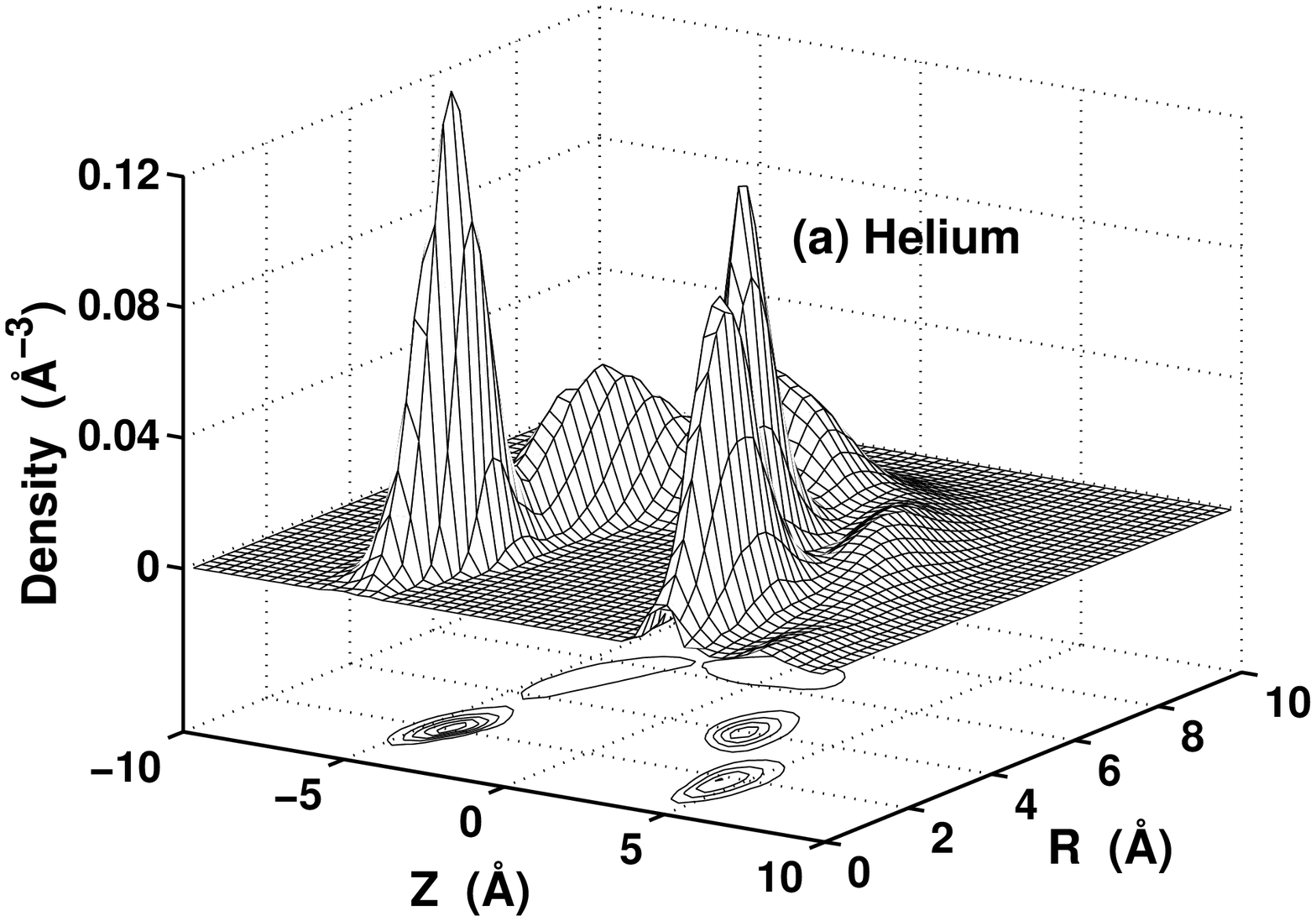}}
\centerline{\includegraphics[height=1.8in,width=2.8in]{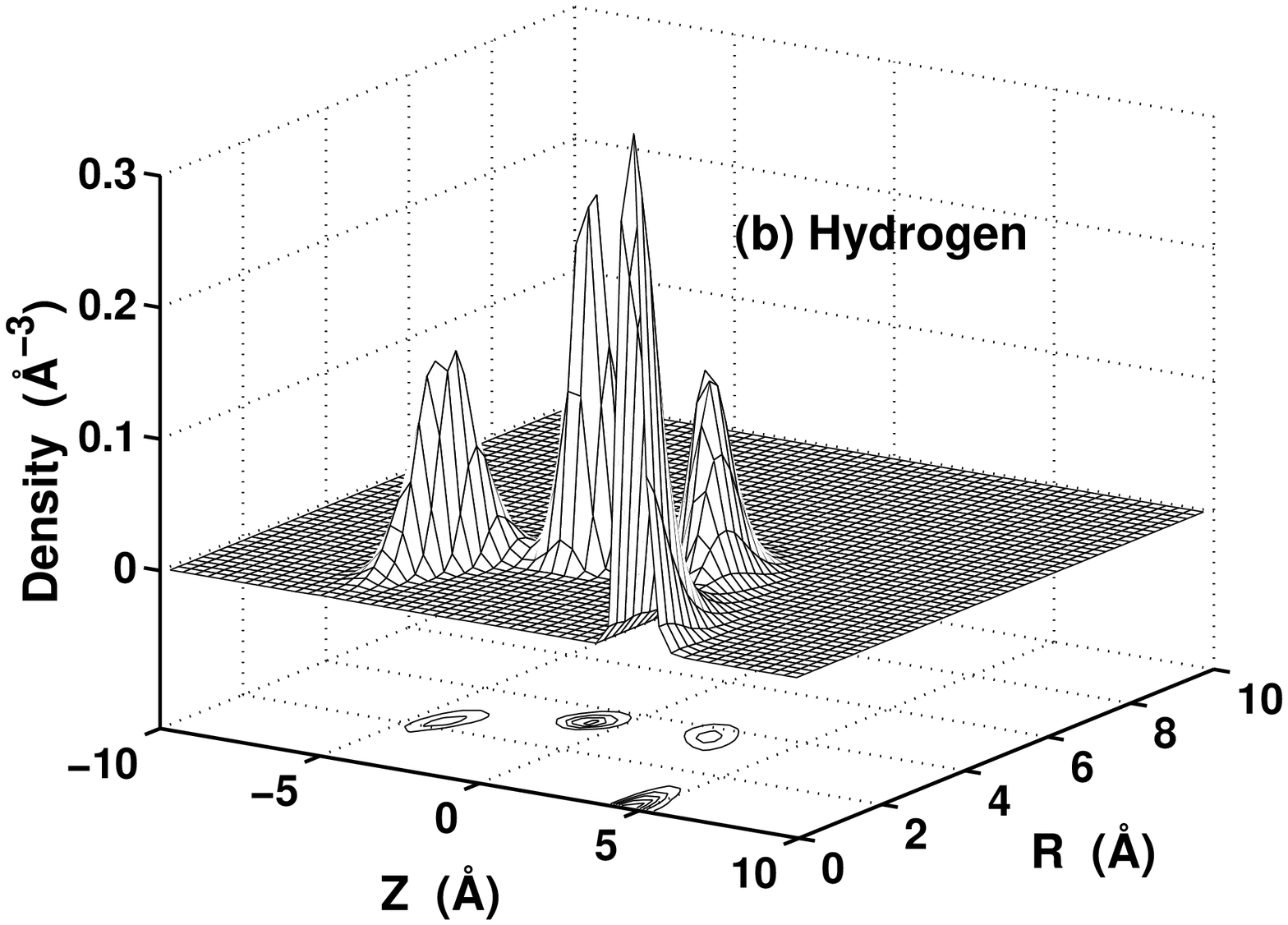}\includegraphics[height=1.8in,width=2.8in]{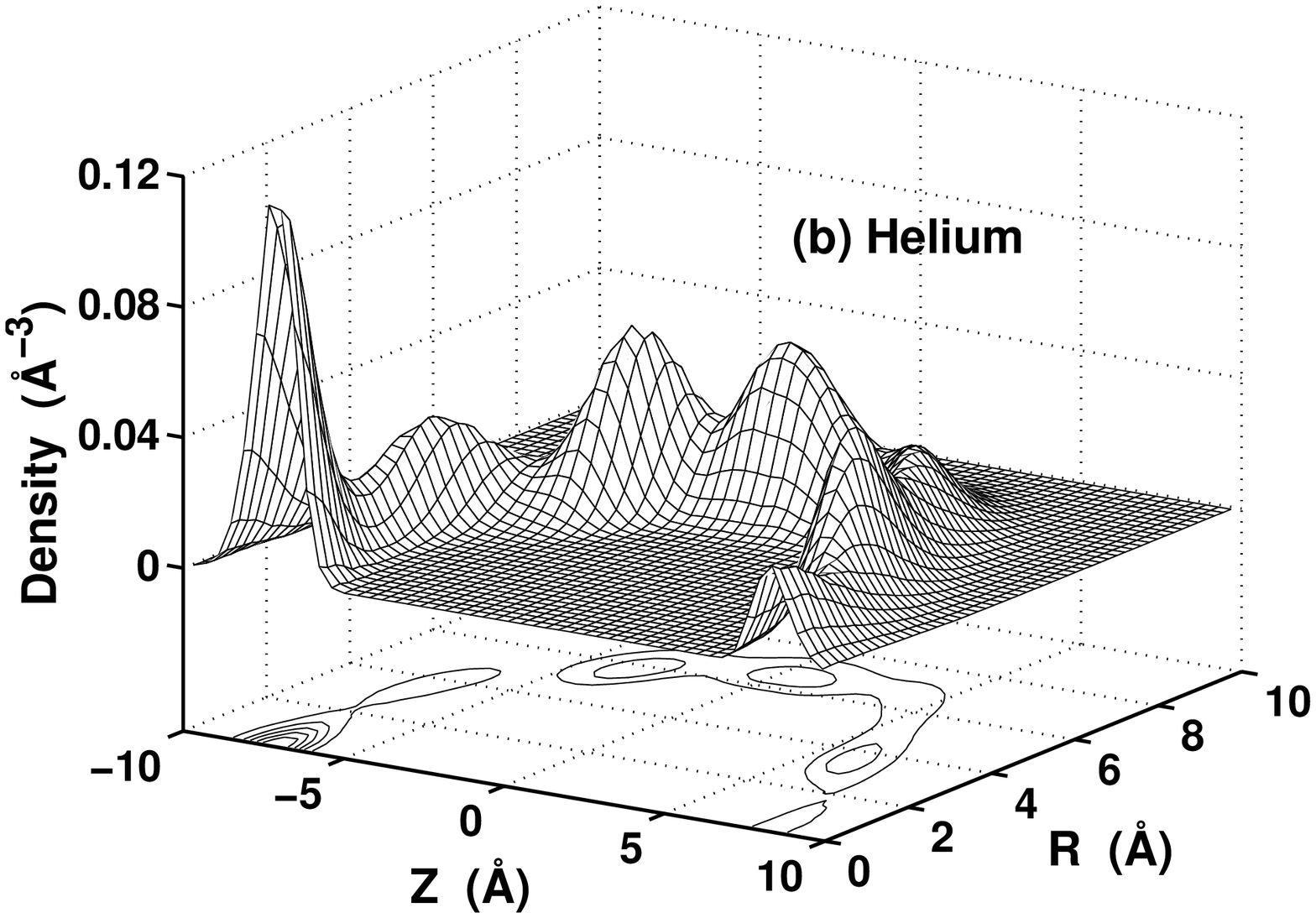}}
\caption{H$_2$ and $^4$He density distributions in OCS(H$_2$)$_M$ complexes 
inside a droplet of 39 $^4$He atoms at T=$0.3125$~K. 
Upper panels, (a), correspond to $M=6$ and lower panels, (b), to $M=17$.
The origin is set at the center of mass of the OCS,
$Z$ is the coordinate along the OCS molecular axis 
with $+Z$ on the sulfur side, 
and $R$ the radial distance from the axis.
}
\label{fig:h2-he.den}
\end{figure}

Figure 2(b) shows the hydrogen and helium distributions for 
the $M=17$ complex embedded in $^4$He$_{39}$.  
The hydrogen is seen to expel all $^4$He atoms from
the immediate vicinity of the OCS molecule and to thus constitute exactly one
solvation layer when the $M=17$ complex is inside
a helium cluster. 
This hydrogen solvation layer consists of
four annular bands around OCS and changes very little 
upon addition of helium.
We showed in Ref.~\onlinecite{kwon02} that
this confined system of hydrogen molecules possesses a transition to 
an anisotropic superfluid state below T$\sim 0.3$~K.
The parallel component $f^s_{\parallel}$, which represents
the superfluid response to rotation around an axis parallel
to the OCS molecular axis, sharply increases to unity at T$ =0.15$~K
from $\sim 0.18$ at T$ \sim 0.3$~K.
The perpendicular
component $f^s_{\perp}$ shows a small increase as the temperature drops below
T$ \sim 0.3$~K but remains much weaker than $f^s_{\parallel}$.
Further analysis showed that a sudden increase in $f^s_{\parallel}$
at the lower temperature is primarily due to the exchange coupling 
among H$_2$ molecules
in different bands.

\section{SPECTROSCOPIC IMPLICATIONS}
The different properties of the $M=1$ and $M=17$ clusters result in 
very different spectroscopic features in helium droplets.  
The $M=1$ cluster behaves as a rigid complex and its spectroscopy can be 
quantitatively interpreted with the assumption of rigid coupling of the H$_2$ 
motion to the OCS motion, together with a contribution from rigidly coupled 
helium non-superfluid density in both the first 
and second solvation shells.\cite{kwon03}  
In contrast, the $M=17$ cluster, which should possess symmetric top symmetry 
and show a $Q$-branch (corresponding to $\Delta J=0$ transitions) in its 
rotational spectrum if analogously rigidly coupled, actually shows 
no $Q$-branch in its infrared spectrum at T$ =0.15$~K.\cite{grebenev00}  
This is rationalized in terms of the high parallel superfluid response 
of the completed hydrogen solvation layer
below $T \sim 0.3$ K, which  
implies that there can be no angular momentum excitation parallel
to the OCS molecular axis in the hydrogen layer, and hence no possibility of a 
$Q$-branch.\cite{kwon02}

The $M=5$ and $6$ clusters lie inbetween these two extremes of minimal 
and complete hydrogen solvation.  Prediction of their infrared spectroscopy 
in helium requires careful analysis of the extent of rigidity and 
the role of exchange permutations in the hydrogen band.  
Further work is required to determine whether the linear response 
quantification of rigidity for a superfluid is appropriate to a single 
annular density band, and how to evaluate the corresponding quantum response 
quantifier. We note that the high symmetry of the annular density bands 
for the $M=5$ and 6 complexes of OCS(H$_2$)$_M$ seen here can be used 
to justify a simple molecular argument for absence of the $Q$-branch based 
on the symmetry of an equally-spaced ring 
of five or six H$_2$ molecules distributed at the waist of the OCS 
molecule.\cite{grebenev02}  A more complete analysis of the quantum 
response will provide critical insight into the onset of nanoscale superfluidity
in these molecular hydrogen solvation clusters.
Since the presence of H$_2$ molecules can induce a nonsuperfluid density
in the second helium solvation shell as well as in the first shell,\cite{kwon03}
further calculations on bigger sizes of helium clusters are necessary
for complete analysis of the local nonsuperfluid helium density
and hence for quantitative estimation of the rotational constants
of the OCS(H$_2$)$_M$ complexes inside helium droplets.

\section{Acknowledgments}
This work has been supported by 
Konkuk University (the 2002 research fund to YK)
and by the Chemistry
Division of the National Science Foundation (grant CHE-0107541 to KBW).

\end{document}